\documentstyle[aps]{revtex}
\begin{document}
\draft
\title{
Mean-field and fluctuation analyses 
of a forced turbulence 
simulated by the lattice Boltzmann method 
}
\author{
W. Sakikawa and O. Narikiyo
}
\address{
Department of Physics, 
Kyushu University, 
Fukuoka 810-8560, 
Japan
}
\date{
July 16, 2003
}
\maketitle
\begin{abstract}
On the basis of the lattice Boltzmann method 
we have done a numerical experiment of a forced turbulence 
in real space and time. 
Our new findings are summarized into two points. 
First 
in the analysis of the mean-field behavior of the velocity field 
using the exit-time statistics  
we have verified Kolmogorov's scaling and Taylor's hypothesis 
for the first time 
in the simulation for the Navier-Stokes equation. 
Second 
in the analysis of the intermittent velocity fluctuations 
using a non-equilibrium probability distribution function 
and the wavelet denoising 
we have clarified that the coherent vortices sustain 
the power-law velocity correlation in the non-equilibrium state. 
\end{abstract}
\vskip 18pt
\section{Introduction}
\label{intro}

Understanding turbulence has not been attained over millenniums. 
However, new approaches to turbulence\cite{Frisch} 
have been invented successively. 

Recently turbulence data in nature and laboratories 
are analyzed from new view points. 
One\cite{exit-time-stat} is based on the exit-time statistics 
and discusses Kolmogorov's scaling\cite{Frisch}. 
The other\cite{PDF-space} 
on the basis of the R\'enyi-Tsallis statistics\cite{RT1,RT2} 
discusses the probability distribution functions (PDF). 

Our study starts from the application of the above two schemes 
to our numerical simulation data 
and aims to obtain some new insight of the turbulence. 

Before reporting our study of the turbulence 
we briefly review here 
our strategy approaching intermittency problems 
in other non-equilibrium systems. 
In the numerical study 
of the critical spin-state of the Ising model\cite{SN1} 
we have observed 
that the degree of the non-Gaussianity of the R\'enyi-Tsallis PDF 
represents that of the non-equilibrium. 
On the other hand, 
in the numerical study of a supercooled liquid\cite{SN2} 
near the glass transition 
we have observed 
that the spatial distribution of the coherently moving regions 
is well analyzed by the singularity spectrum of multifractal. 
Since the R\'enyi-Tsallis statistics can describe 
the systems with multifractality, 
the above two observations are closely related. 

In these two systems 
the correlation length of the fluctuations becomes divergently large 
and we can expect the scale-invariance. 
A scale-invariant system has multifractal nature. 
On the other hand, 
we can also expect the scale-invariance 
in the turbulent systems at high Reynolds number\cite{Frisch}. 
Thus we adopt the same strategy 
to intermittency problem in turbulence. 

The next section describes our simulation method. 
In the following sections the simulation data are analyzed 
at the mean-field level at first 
and subsequently taking fluctuations into account. 
In the final section a brief summary is given. 


\section{Lattice Boltzmann simulation}
\label{LBsimulation}

We simulate a forced turbulence in real space and time 
and adopt the lattice Boltzmann method\cite{LB2,LB1} 
as one of the easiest ways for such a purpose. 

The distribution function $f_i({\vec r},t)$ 
of the particle with the velocity ${\vec c}_i$ 
at position ${\vec r}$ and time $t$ 
obeys the lattice Boltzmann equation\cite{LB2,LB1} 
\begin{equation}
f_i({\vec r}+{\vec c}_i,t+1) - f_i({\vec r},t)
= - \omega [ f_i({\vec r},t) - f_i^0({\vec r},t) ],
\label{f}
\end{equation}
where we have adopted the Bhatnagar-Gross-Krook (BGK) model 
for the collision term 
and the time step has been chosen as unity. 
The position ${\vec r}$ of the particle 
is restricted on the cubic lattice 
and the lattice spacing is chosen as unity. 
The local equilibrium distribution $f_i^0({\vec r},t)$ is assumed 
to be given as 
\begin{equation}
f_i^0({\vec r},t)
= w_i \rho [ 1 + 3 ({\vec c}_i\cdot{\vec u}) 
               + {9 \over 2} ({\vec c}_i\cdot{\vec u})^2 
               - {3 \over 2} ({\vec u}\cdot{\vec u}) ].
\label{f0}
\end{equation}
The density $\rho$ and the velocity ${\vec u}$ of the fluid 
are given by the sum of the contributions of the particles as 
$\rho = \sum_i f_i({\vec r},t)$ and 
$\rho {\vec u} = \sum_i f_i({\vec r},t) {\vec c}_i$, respectively. 
Using the multiscale analysis 
the Navier-Stokes equation, 
which is supposed to be able to describe turbulence, 
is derived from the present BGK lattice Boltzmann equation 
and the corresponding viscosity $\nu = (1/\omega - 1/2)/3$. 

In the following 
we use a 15-velocity model and ${\vec c}_i$ is chosen as 
${\vec c}_1 = (0,0,0)$, 
${\vec c}_i = (\pm1,0,0),(0,\pm1,0),(0,0,\pm1)$ 
for $i=2,3,\cdot\cdot\cdot,7$ 
and 
${\vec c}_i = (\pm1,\pm1,\pm1)$ for $i=8,9,\cdot\cdot\cdot,15$. 
Then the weight factor $w_i$ for the local equilibrium distribution 
is determined as 
$w_1 = 2/9$, 
$w_i = 1/9$ for $i=2,3,\cdot\cdot\cdot,7$ and 
$w_i = 1/72$ for $i=8,9,\cdot\cdot\cdot,15$. 
The relaxation frequency $\omega$ is chosen as $\omega=1.94$. 

In order to simulate forced turbulence 
we add the forcing term ${\vec g}_i\cdot{\vec F}({\vec r},t)$ 
due to an applied body force ${\vec F}({\vec r},t)$ 
to the right hand side of Eq.\ (\ref{f}). 
In order to reproduce the Navier-Stokes equation 
we should choose as ${\vec g}_i = {\vec c}_i/10$ 
for the present 15-velocity model\cite{LB2}. 
For simplicity we apply a solenoidal force in $y$-direction\cite{LB1}, 
${\vec F}=(0,F_y(z,t),0)$, and $F_y(z,t)$ is a Gaussian white noise 
whose variance $\sigma_F$ is a function of the lattice coordinate 
in $z$-direction, 
$\sigma_F = |0.01 \times \sin(2\pi k_f z / L)|$, 
where $L$ is the linear dimension 
of the simulation region of cubic box. 
We have chosen as $L = 200$ so that $1 \leq x,y,z \leq 200$. 
We have adopted the periodic boundary condition. 
Although we have chosen as $k_f = 4$, 
the results of the simulation is insensitive to the choice. 

In the initial state of the simulation 
the fluid density is uniform, $\rho = 1$, and 
the spatial distribution of the fluid velocity ${\vec u}$ 
is chosen to be random. 

In Fig.\ \ref{fig1} 
we show a snapshot of the vorticity field, 
$\vec \omega = \nabla \times \vec u$, 
in order to visualize a turbulent structure. 
We see an inhomogeneous distribution of the vortices. 

The Reynolds number $Re$ for our simulation\cite{LB2,LB1} is 
estimated as $Re = \sqrt{ \langle {\vec u}^2 \rangle } L / 2\pi\nu$ 
and time-dependent as shown in Fig.\ \ref{fig2}. 
Here $\langle \cdot\cdot\cdot \rangle$ represents the spatial average. 
In our simulation 
we can realize a turbulent state for relatively small Reynolds number, 
since we add random force. 


\section{Mean-field behavior}
\label{MF}

In this section 
we discuss the mean-field aspects of the velocity field. 

Turbulence is one of the typical phenomena 
with multiscale motions. 
Each phenomena of a scale strongly couples with 
all the other scales of turbulent motion. 
In order to analyze such a system 
a scale-dependent entropy, the so-called $\epsilon$-entropy, 
works well\cite{exit-time-stat}. 
For example, the time series of the velocity fluctuation in turbulence 
leads to a non-trivial scaling relation of the $\epsilon$-entropy, 
$ h(\epsilon) \propto \epsilon^{-3} $, 
expected from Kolmogorov's scaling\cite{exit-time-stat}. 
The existing experimental data are consistent 
with this scaling\cite{exit-time-stat}. 
In this section we try to show the consistency of the scaling 
with our numerical experiment simulating the Navier-Stokes equation. 

We focus our attention to the time series of 
the $y$-component $u_y({\vec r},t)$ of 
the fluid velocity ${\vec u}({\vec r},t)$ 
as shown in Fig.\ \ref{fig3}, 
since our system is anisotropic due to the forcing. 
This turbulent signal is characterized by the exit-time 
$\tau(\epsilon;{\vec r})$ at position ${\vec r}$. 
At the exit-time $t=\tau(\epsilon;{\vec r})$ 
the first exit satisfying the condition, 
$|u_y({\vec r},t_0+\tau) - u_y({\vec r},t_0)| > \epsilon/2$, 
occurs. 
Here the time $\tau$ is measured from $t_0$ and $t_0=1000$ in our simulation 
regarding the states at the first 1000 time-steps as transitional. 

The mean of the exit-time, 
$\tau(\epsilon) \equiv \langle \tau(\epsilon;{\vec r}) \rangle$, 
is related to the $\epsilon$-entropy, $h(\epsilon)$, 
as\cite{exit-time-stat} 
\begin{equation}
h(\epsilon) \propto {1 \over \tau(\epsilon)}. 
\label{h}
\end{equation}
In the region 
where Kolmogorov's scaling, the 4/5 law\cite{Frisch,exit-time-stat}, holds 
the velocity difference between two points behaves as 
\begin{equation}
\langle|{\vec u}({\vec r}+{\vec R},t_0)-{\vec u}({\vec r},t_0)|\rangle 
     \propto |{\vec R}|^{1/3} . 
\label{space-scaling}
\end{equation}
Using Taylor's hypothesis\cite{Frisch,exit-time-stat}, we obtain 
\begin{equation}
\langle|{\vec u}({\vec r},t_0+\tau)-{\vec u}({\vec r},t_0)|\rangle 
     \propto \tau^{1/3},
\label{time-scaling}
\end{equation}
as a mean-field description. 
In spite of the absence of the global flow in our simulation 
Taylor's hypothesis is valid in a local sense\cite{JPV} 
as shown in the following. 
Namely in the mean-field description 
the velocity difference in temporal and spatial directions 
have the same fractal scaling exponent.
With Eq.\ (\ref{time-scaling}) the definition of the exit-time leads to 
$\tau(\epsilon)^{1/3} \propto \epsilon$. 
From Eq.\ (\ref{h}) we can conclude 
that $h(\epsilon) \propto \epsilon^{-3}$ 
in the above mean-field description\cite{exit-time-stat}. 

In accordance with the above scaling analysis 
our simulation data\cite{SN3} for $1/\tau(\epsilon)$ 
shown in Fig.\ \ref{fig4} 
have the scaling region for $\epsilon > \epsilon_0$ 
where $\epsilon_0 \sim 0.04$. 
Thus by our simulation we have established 
that $h(\epsilon) \propto \epsilon^{-3}$. 
For $\epsilon \ll \epsilon_0$ 
the exit in the simulation occurs within one time step 
and the curve saturates at small $\epsilon$. 
These behaviors are consistent 
with the experimantal data\cite{exit-time-stat}. 

In Fig.\ \ref{fig5} 
the result of the similar analysis in spatial direction is shown. 
In place of $\tau(\epsilon)$ we measure $r(\epsilon)$ 
where at the exit-length $r=r(\epsilon;{\vec r})$ 
the first exit satisfying the condition, 
$|u_y(x+r,y,z,t) - u_y(x,y,z,t)| > \epsilon/2$, 
occurs at a time $t$. 
Here $r(\epsilon) = \langle r(\epsilon;{\vec r}) \rangle$. 
Because of the smallness of the number of the available data 
in space direction, 200, in comparison with that in time direction, 7000, 
the scaling region is narrow. 
Since $1/\tau(\epsilon)$and $1/r(\epsilon)$ 
have the same scaling exponent, 
Taylor's hypothesis is confirmed to be valid. 

In conclusion we have confirmed 
Kolmogorov's scaling and Taylor's hypothesis at the same time. 
In contrast to the evaluation of the energy spectrum, 
which is usually employed for testing Kolmogorov's scaling 
and determined by the two-point correlation function, 
Kolmogorov's scaling is easily observed in the $\epsilon$-entropy, 
since it is a mean-field description and fluctuations are averaged out. 
Such a mean-field scaling is a unifractal description and 
fluctuations can be taken into account in a multifractal analysis 
as shown in the following. 


\section{Fluctuations: qualitative observation}
\label{quality}

We have discussed the mean-field aspects of the velocity field 
in the preceding section. 
Next we discuss the fluctuations of the velocity field. 
In this section we make a qualitative observation 
using the correlation maps. 

In Fig.\ \ref{fig6} we show the space correlation. 
The correlation decays as the length scale is increased. 
In Fig.\ \ref{fig7} we show the time correlation. 
The correlation decays as the time scale is increased. 
At small length or time scale 
the strong correlation leads to a non-equilibrium state 
which is identified as the coherent vortex in the following section. 

We see a multiscale structure in the correlation map, 
Fig.\ \ref{fig8}, 
which is one of the characteristics of turbulence 
neither periodic nor random. 


\section{Fluctuations: non-equilibrium nature}
\label{non-equilibrium-nature}

We have discussed the qualitative aspects of the velocity fluctuation 
in the preceding section. 
Next we try to quantify the fluctuation 
by two measures. 
One is the non-equilibrium parameter used in this section and 
the other is the singularity spectrum in the next section. 

In order to analyze the distribution of the fluctuation 
we use the R\'enyi-Tsallis PDF\cite{RT1,RT2} 
\begin{equation}
P(X) = P_0 \cdot [ 1 + {q-1 \over 2\sigma_X^2} X^2 ]^{1 \over 1-q}, 
\label{PDF-RT}
\end{equation}
for a variable $X$. 
Here the parameter $q$ represents 
the degree of non-equilibrium\cite{SN1}. 
To be precise, 
the deviation of $q$ from unity 
is the measure for the degree of non-equilibrium, 
since the equilibrium Gaussian distribution 
is expressed by Eq.\ (\ref{PDF-RT}) 
in the limit of $q \rightarrow 1$. 

The PDF in Eq.\ (\ref{PDF-RT}) 
is derived from the R\'enyi or Tsallis entropy 
by using the maximum entropy principle\cite{AA}. 
The Tsallis entropy is non-extensive, 
while the R\'enyi entropy is extensive. 
The functional form of the distribution is independent of 
the choice of the entropy, the R\'enyi or Tsallis, 
so that our result is independent of the extensivity of the entropy. 

We show the PDF for the velocity difference 
between two points in Fig.\ \ref{fig9} and 
the vorticity difference in Fig.\ \ref{fig10}. 
The $q$-value of the PDF is length-scale dependent. 
Similarly the PDF for the velocity or vorticity difference 
between two times at a point is time-scale dependent. 
In any case 
the degree of non-equilibrium is larger 
at smaller spatio-temporal scales 
where the correlation survives. 
As shown later on, 
the formation of the coherent vortex structure 
sustains the strong correlation leading to non-Gaussian PDF. 
At larger scales the contribution of the structureless 
incoherent background among vortices dominates. 
This incoherency leads to Gaussian PDF. 
The Gaussian PDF is a parabola in the semi-logarithmic plane 
as Fig.\ \ref{fig9} and Fig.\ \ref{fig10}. 

Although our data in Fig.\ \ref{fig9} 
are not enough to discuss the scaling between $q$ and $r$ quantitatively, 
we have to derive such a relation in our future study. 
Already a scaling relation between $q$ and $r$ 
has been proposed\cite{Beck} 
under the assumption of a cascade picture. 
On the other hand, another scheme 
with a $r$-independent $q$-value 
has been proposed\cite{AA} 
under the assumption of another cascade picture. 
In contrast to these two approaches 
our analysis has nothing to do with cascades. 

In Fig.\ \ref{fig11} 
we show the PDF for the vorticity itself 
taken from a snapshot of the vorticity field. 
This PDF is also non-Gaussian. 
The Gaussian components can be removed from the snapshot 
by using the wavelet denoising technique\cite{FSK}. 
It can be seen from the coincidence 
between the left and right panels in Fig.\ \ref{fig12} 
that the non-Gaussian components corresponds to the vortices. 
Here the Gaussian component is filtered out 
by the wavelet denoising in the right panel. 
This visualization is not new, 
since it has already been done 
in the study of the coherent vortex simulation\cite{FSK}. 
Our new contribution is to clarify the origin of 
the scale-dependent $q$-value 
in terms of the coherent vortex structure. 
The strong correlation leading to the non-Gaussian PDF 
at smaller scales is sustained in the inner region of the vortices, 
while the PDF at larger scales than the vortex size tends to Gaussian. 
In other words the inner region is coherent 
and the outer is incoherent. 


\section{Fluctuations: multifractal nature}
\label{multifractal-nature}

We have discussed the local nature of the vortex 
as the elemntary excitation in the preceding section. 
In this section we discuss the spatial distribution 
of the vortices in a global point of view. 

First we make a qualitative observation. 
Already as seen in Fig.\ \ref{fig12} 
the spatial distribution of the coherent vortices is intermittent. 
In Fig.\ \ref{fig13} 
we also see a intermittent behavior 
for the derivative of the vorticity. 
Similar intermittent behavior is observed 
in experimental data\cite{MSKF}. 

Next we quantify the intermittency 
by introducing the measure, $\mu_{k}(l)$, defined as the probability 
of finding a coherent vortex in the $k$-th cubic box 
of the linear dimension $l$ in a snapshot of the vorticity field. 
The total number of the boxes $N_L = (L/l)\times(L/l)$ 
where $L$ is the linear dimension of the simulation region. 
The measure behaves as 
\begin{equation}
\mu_{k}(l) \propto l^{\alpha_k},  
\end{equation}
for small $l/L$. 
The number density $N(\alpha)$ for the exponent $\alpha_k$ is 
defined as 
\begin{equation}
N(\alpha) \equiv \sum_k \delta(\alpha - \alpha_k) 
          \propto l^{-f(\alpha)}.  
\end{equation}
This relation defines the singularity spectrum, 
$f(\alpha)$, which is the key quantity to qualify multifractal systems. 
Using the normalized $q$-th moment $\mu_{k}(q,l)$ 
of the measure $\mu_{k}(l)$, 
\begin{equation}
\mu_{k}(q,l) = \{\mu_{k}(l)\}^{q} 
             / \sum_{k'=1}^{N_L} \{\mu_{k'}(l)\}^{q}, 
\end{equation}
the sigularity spectrum $f(\alpha)$ is obtained 
by the following formulae,\cite{Schreiber} 
\begin{equation}
\alpha(q) = \sum_{k=1}^{N_L} \mu_{k}(q,l) \ln \mu_{k}(l) / \ln (l/L),  
\end{equation}
and 
\begin{equation}
f(\alpha(q)) = \sum_{k=1}^{N_L} \mu_{k}(q,l) \ln \mu_{k}(q,l) / \ln (l/L).
\end{equation}

In Fig.\ \ref{fig14} 
we show the singularity spectrum 
for the coherent vortex. 
For simplicity we have used the definition $N_k/N_L = l^{\alpha_k}$ 
where $N_k$ is the number of lattice points 
satisfying the condition $\omega_z > \langle \omega_z \rangle$. 
The spectrum is similar to the typical one 
measured for the energy-dissipation rate 
by experiments\cite{MSKF} or simulations\cite{HY}. 

In the last part of this section 
we give some speculations. 
As seen in our previous study\cite{SN2} 
the width of the singularity spectrum 
depends on the degree of intermittency 
so that we expect some Reynolds-number dependence 
of the spectrum. 
It has been discussed by many authors\cite{Sreenivasan} 
and should be claryfied in future systematic study. 
In our present study 
the density of the vortex is dilute 
as seen in Fig.\ \ref{fig12}. 
Thus the singularity spectrum in Fig.\ \ref{fig14} 
describes the spatial distribution of relatively free vortices. 
As the Reynolds number is increased 
the density increases\cite{KM}. 
In this case 
the interaction among vortices becomes important and  
the correlation length of the vorticity fluctuation becomes large. 
While we can observe the vortex 
only as an individual elementary excitation 
in our numerical experiment, 
some collective excitation is expected to dominate 
at higher Reynolds number. 
In the limit of divergently large Reynolds number 
the correlation length becomes divergently large 
so that we can expect full scale-invariance. 
We can find a resemblance to the case of the scaling theory\cite{deGennes} 
in polymers where an ideal scaling relation is realized for dense solutions 
where polymers are strongly entangled. 
In the limit of high Reynolds number 
each boxes counting the coherent vortex 
will be filled by almost equal number of vortices 
so that intermittency will disappear and 
unifractal Kolmogorov's scaling will prevail. 


\section{Summary}
\label{summary}

Numerical simulation data, in real space and time, 
for a forced turbulence 
on the basis of the lattice Boltzmann method 
have been analyzed by unifractal and multifractal schemes. 

Our new findings are summarized into two points. 
First in the unifractal analysis using the exit-time statistics 
we have verified Kolmogorov's scaling and Taylor's hypothesis 
at the same time. 
Second in the analysis 
using the R\'enyi-Tsallis PDF and the wavelet denoising 
we have clarified that the coherent vortices sustain 
the power-law velocity correlation in the non-equilibrium state. 

Finally in the multifractal analysis 
it is clarified 
that the intermittent distribution 
of the coherent vortices in space-time 
is described as a multifractal. 

This work was supported in part 
by a Grand-in-Aid for Scientific Research 
from the Ministry of Education, Culture, Sports, Science 
and Technology of Japan. 


\vskip 15pt
\begin{figure}
\caption{
The snapshot of the $z$-component of the vorticity, $\omega_z$, 
in the $xy$-plane at $z=100$ and $t=4000$. 
}
\label{fig1}
\end{figure}

\vskip -15pt

\begin{figure}
\caption{
The Reynolds number $Re$ as a function of time $t$. 
}
\label{fig2}
\end{figure}

\vskip -15pt

\begin{figure}
\caption{
The $y$-component of the fluid velocity $u_y$ 
as a function of time $t$ at ${\vec r}=(100,100,100)$. 
}
\label{fig3}
\end{figure}

\vskip -15pt

\begin{figure}
\caption{
The inverse of the mean exit-time $\tau(\epsilon)$. 
The exits are measured at $200^3$ lattice points for $1000 \le t \le 8000$. 
Here the dots and straight line represent the simulation data 
and $\epsilon^{-3}$ law, respectively. 
}
\label{fig4}
\end{figure}

\vskip -15pt

\begin{figure}
\caption{
The inverse of the mean exit-length $r(\epsilon)$. 
The exits are measured at $200^3$ lattice points at $t=6000$. 
Here the dots and straight line represent the simulation data 
and $\epsilon^{-3}$ law, respectively. 
}
\label{fig5}
\end{figure}

\vskip -15pt

\begin{figure}
\caption{
The space correlation of the velocity fluctuation at $t=4000$. 
The horizontal and vertical axes represent $u_y(x,y,z,t)$ and 
$u_y(x,y+y_{\rm corr},z,t)$, respectively. 
The space difference $y_{\rm corr}$ is taken as 
$y_{\rm corr}=1$ for the left, 
$y_{\rm corr}=4$ for the center and 
$y_{\rm corr}=30$ for the right. 
}
\label{fig6}
\end{figure}

\vskip -15pt

\begin{figure}
\caption{
The time correlation of the velocity fluctuation 
at $\vec r =(100,100,100)$. 
The horizontal and vertical axes represent $u_y(\vec r,t)$ and 
$u_y(\vec r,t+t_{\rm corr})$, respectively for $1000 \le t \le 8000$. 
The time difference $t_{\rm corr}$ is taken as 
$t_{\rm corr}=10$ for the left, 
$t_{\rm corr}=100$ for the center and 
$t_{\rm corr}=500$ for the right. 
}
\label{fig7}
\end{figure}

\vskip -15pt

\begin{figure}
\caption{
The correlation map for the velocity difference between two times 
$t_i$ and $t_j$ 
for $1000 \le t_0+t_i \le 8000$ and $1000 \le t_0+t_j \le 8000$ 
with $t_0=1000$. Namely $0 \le i \le 7000$ (horizontal axis) and 
$0 \le j \le 7000$ (vertical axis). 
The dot is marked at $(i,j)$ if $D(i,j)<\theta$ 
where $D(i,j)=|u_y(\vec r,t_i)-u_y(\vec r,t_j)|$ 
and $\theta=0.2D_{\rm max}$ 
with $D_{\rm max}$ being the maximum of $D(i,j)$. 
The position $\vec r$ is fixed at $(100,100,50)$. 
}
\label{fig8}
\end{figure}

\vskip -15pt

\begin{figure}
\caption{
The unnormalized PDF $P(X)$ for the velocity difference 
$X=|u_y(x,y+\Delta y,z,t)-u_y(x,y,z,t)|$ 
between two points $(x,y+\Delta y,z)$ and $(x,y,z)$ at $t=4000$. 
The difference in $y$-direction is chosen as $\Delta y = 1, 4, 30$. 
The data are fitted by the R\'enyi-Tsallis PDF with 
$q = 1.218, 1.164, 1.100$ for $\Delta y = 1, 4, 30$, respectively. 
}
\label{fig9}
\end{figure}

\vskip -15pt

\begin{figure}
\caption{
The unnormalized PDF $P(X)$ for the vorticity difference 
$X=|\omega_y(x,y+\Delta y,z,t)-\omega_y(x,y,z,t)|$ 
between two points $(x,y+\Delta y,z)$ and $(x,y,z)$ at $t=6000$. 
The difference in $y$-direction is chosen as $\Delta y = 1, 4, 30$. 
The data are fitted by the R\'enyi-Tsallis PDF with 
$q = 1.240, 1.209, 1.179$ for $\Delta y = 1, 4, 30$, respectively. 
}
\label{fig10}
\end{figure}

\vskip -15pt

\begin{figure}
\caption{
The unnormalized PDF $P(X)$ for the vorticity 
$X=|\omega_y(x,y,z,t)|$ at $t=6000$. 
The data are fitted by the R\'enyi-Tsallis PDF with $q = 1.392$. 
}
\label{fig11}
\end{figure}

\vskip -15pt

\begin{figure}
\caption{
The snapshot of the $z$-component of the vorticity, $\omega_z$, 
in the $xy$-plane at $z=100$ and $t=4000$. 
In contrast to Fig.\ \ref{fig1} 
the regions satisfying the condition, 
$\omega_z > \langle \omega_z \rangle + 2\sigma \equiv \Theta$, 
are depicted in the left panel. 
Here $\langle \omega_z \rangle$ is the spatial average 
and $\sigma$ is the standard deviation. 
In the right panel 
the high vorticity regions satisfying the same condition, 
$\tilde \omega_z > \Theta$, are depicted 
where $\tilde \omega_z$ is the non-Gaussian component 
after the wavelet denoising filtering out the Gaussian component. 
By the restriction of the algorithm of the wavelet transform 
we have used only 128$\times$128 lattice points. 
}
\label{fig12}
\end{figure}

\vskip -15pt

\begin{figure}
\caption{
The spatial distribution of the vorticity gradient, 
$\Delta\omega_z = \omega_z(x+1,y,z,t) - \omega_z(x,y,z,t)$, 
in $x$-direction ($1 \le x \le 200$) with $y=z=100$ at $t=6000$. 
}
\label{fig13}
\end{figure}

\vskip -15pt

\begin{figure}
\caption{
The singularity spectrum of the coherent vortex with $l/L = 1/10$. 
The data correspond to $-90 \le q \le 95$. 
}
\label{fig14}
\end{figure}
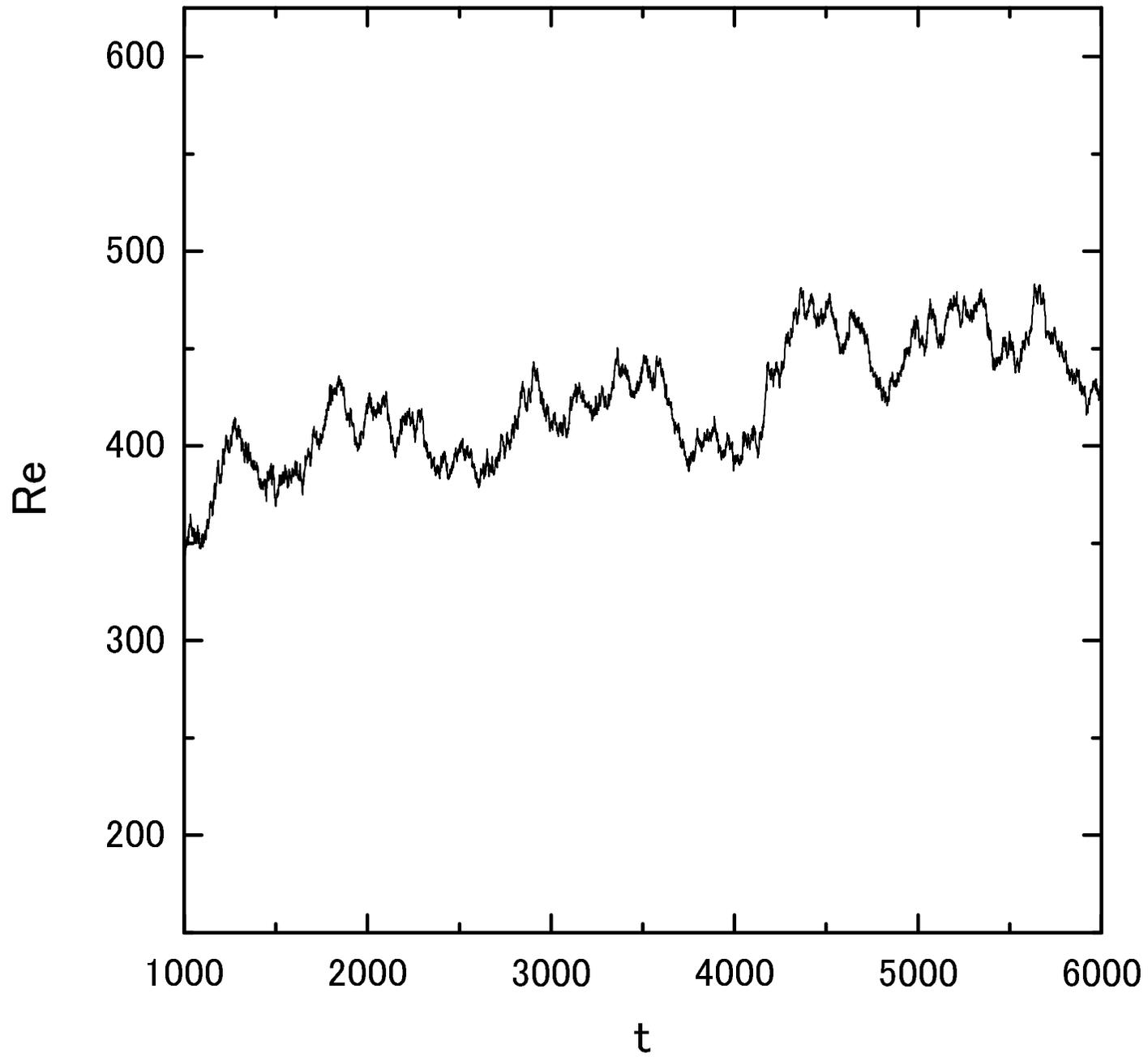
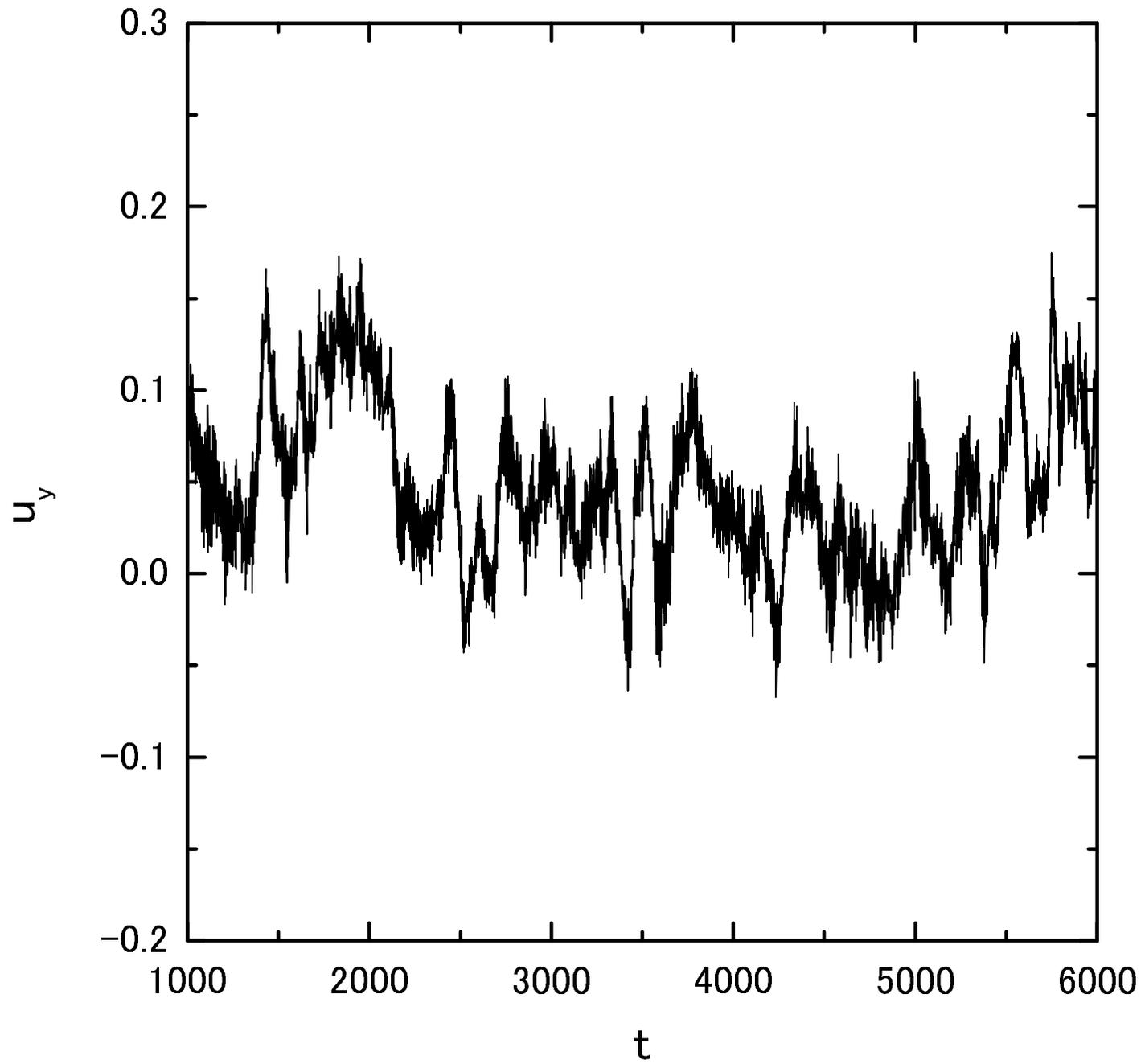
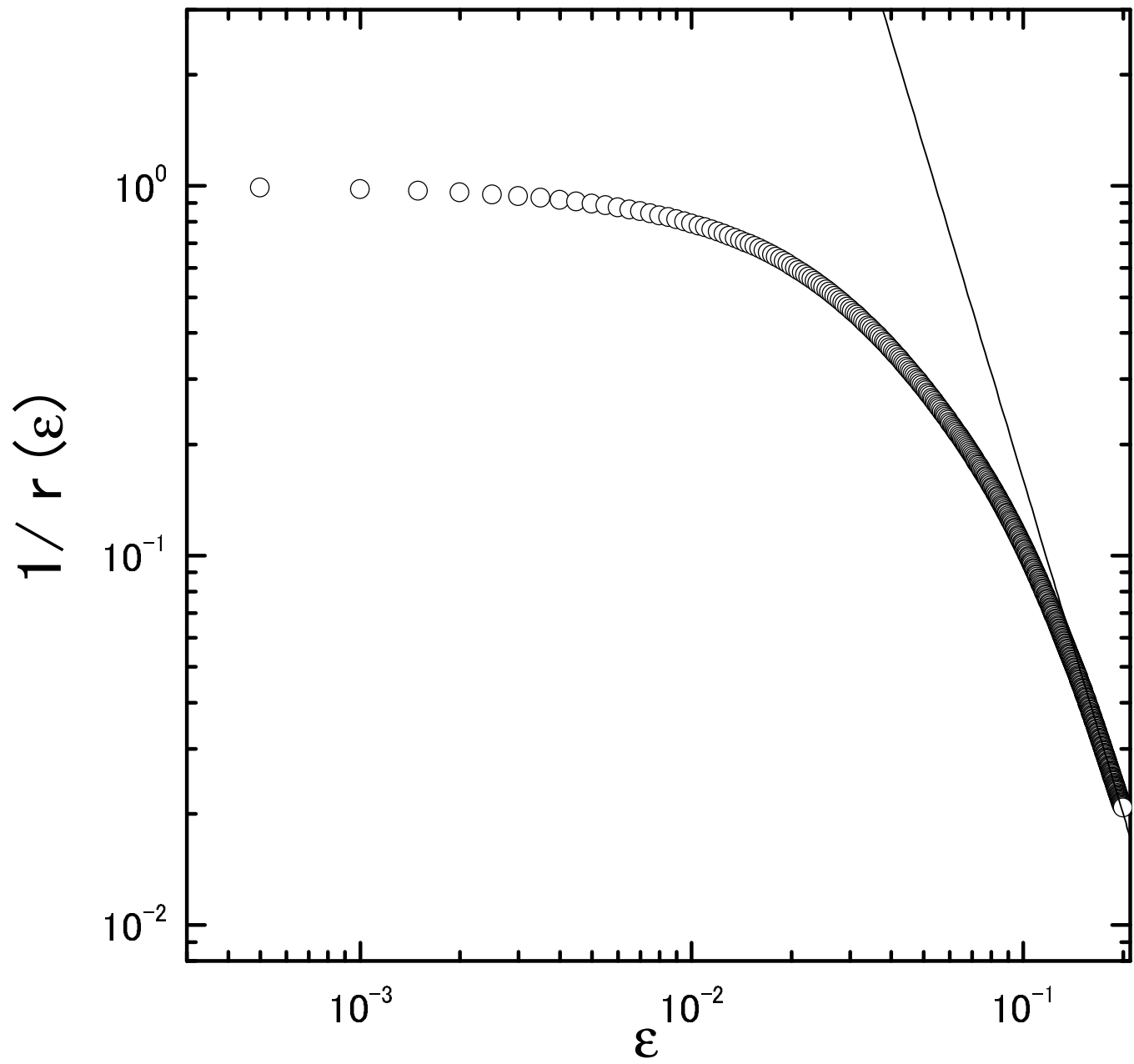
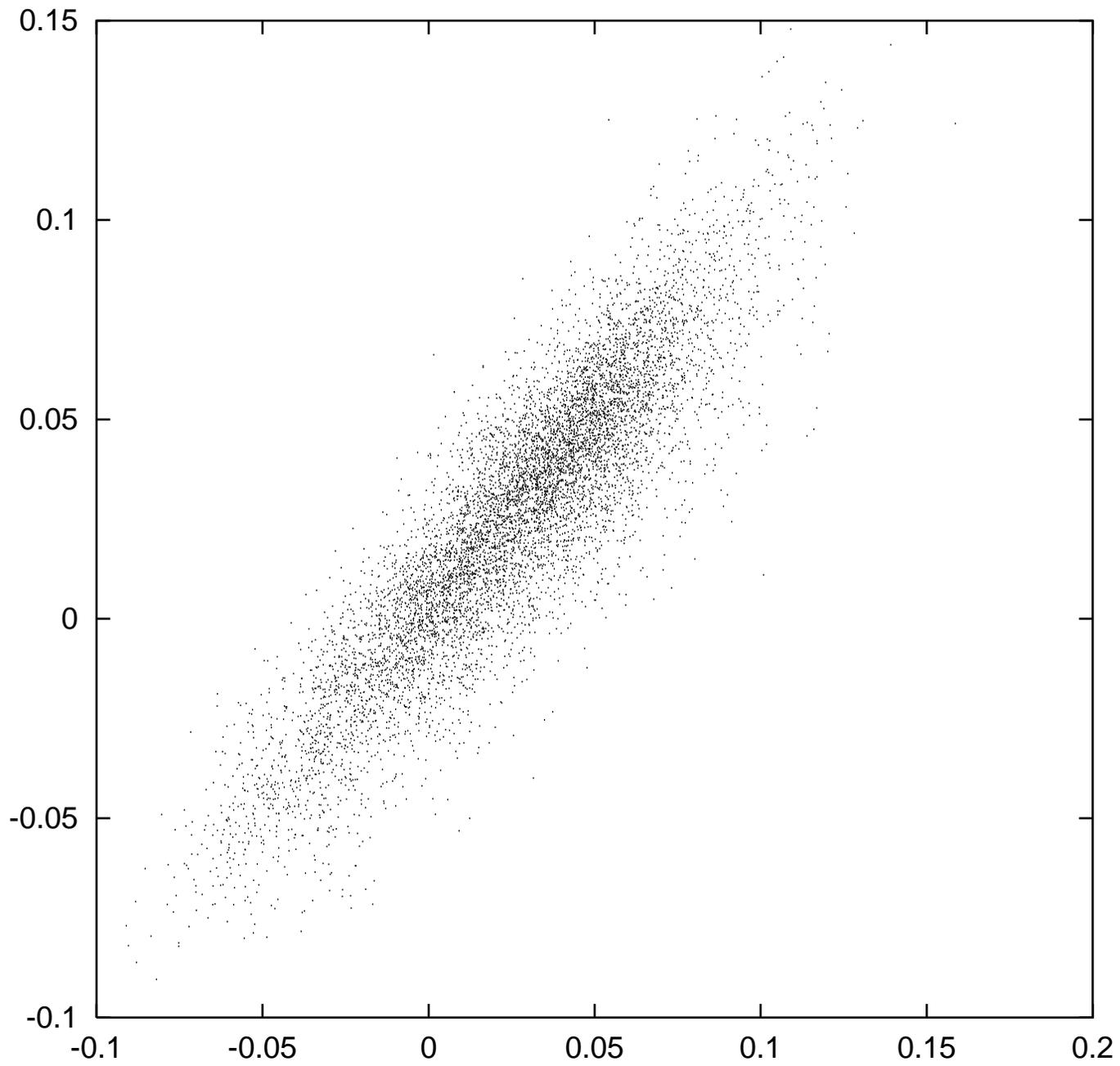
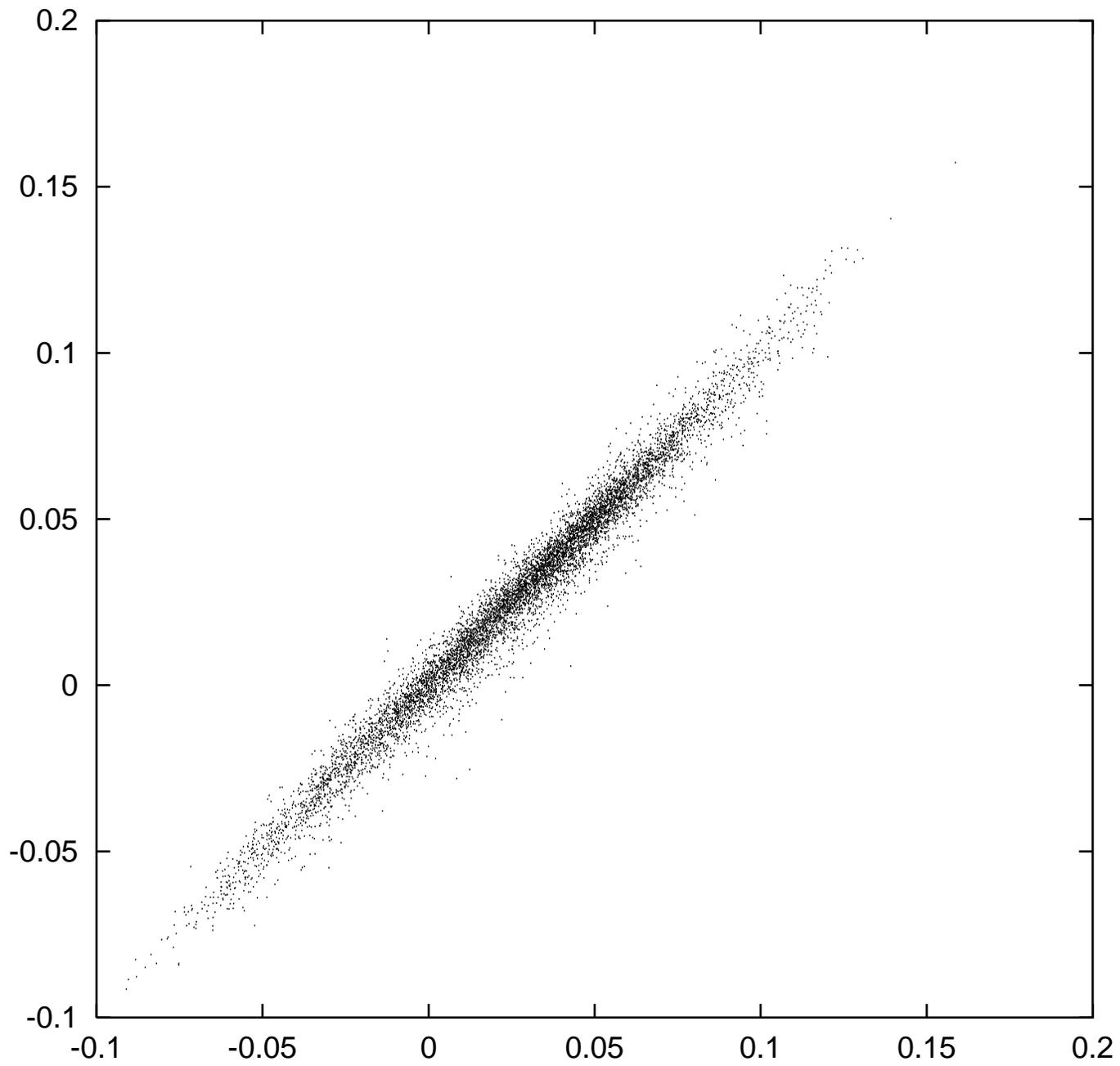
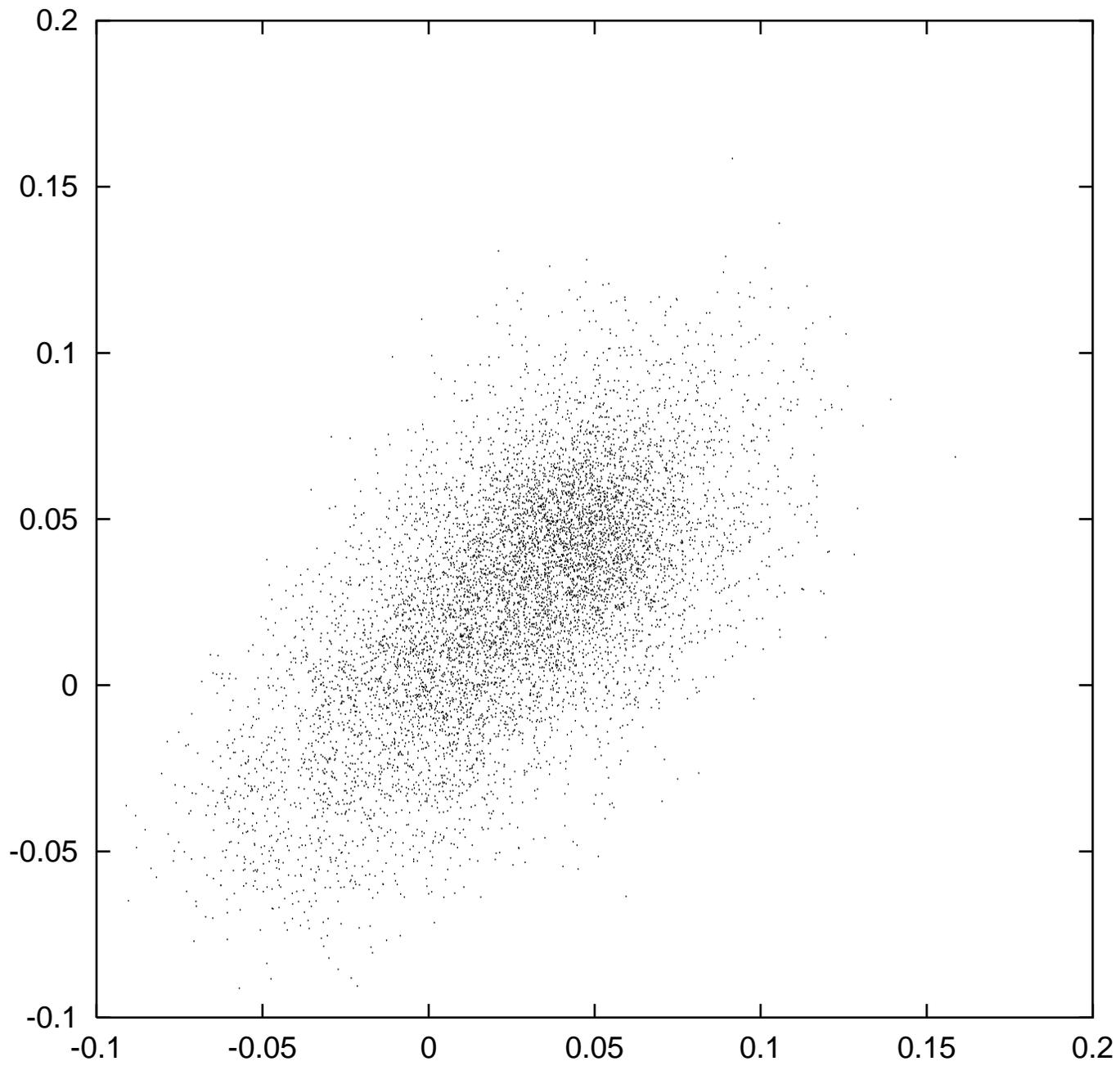
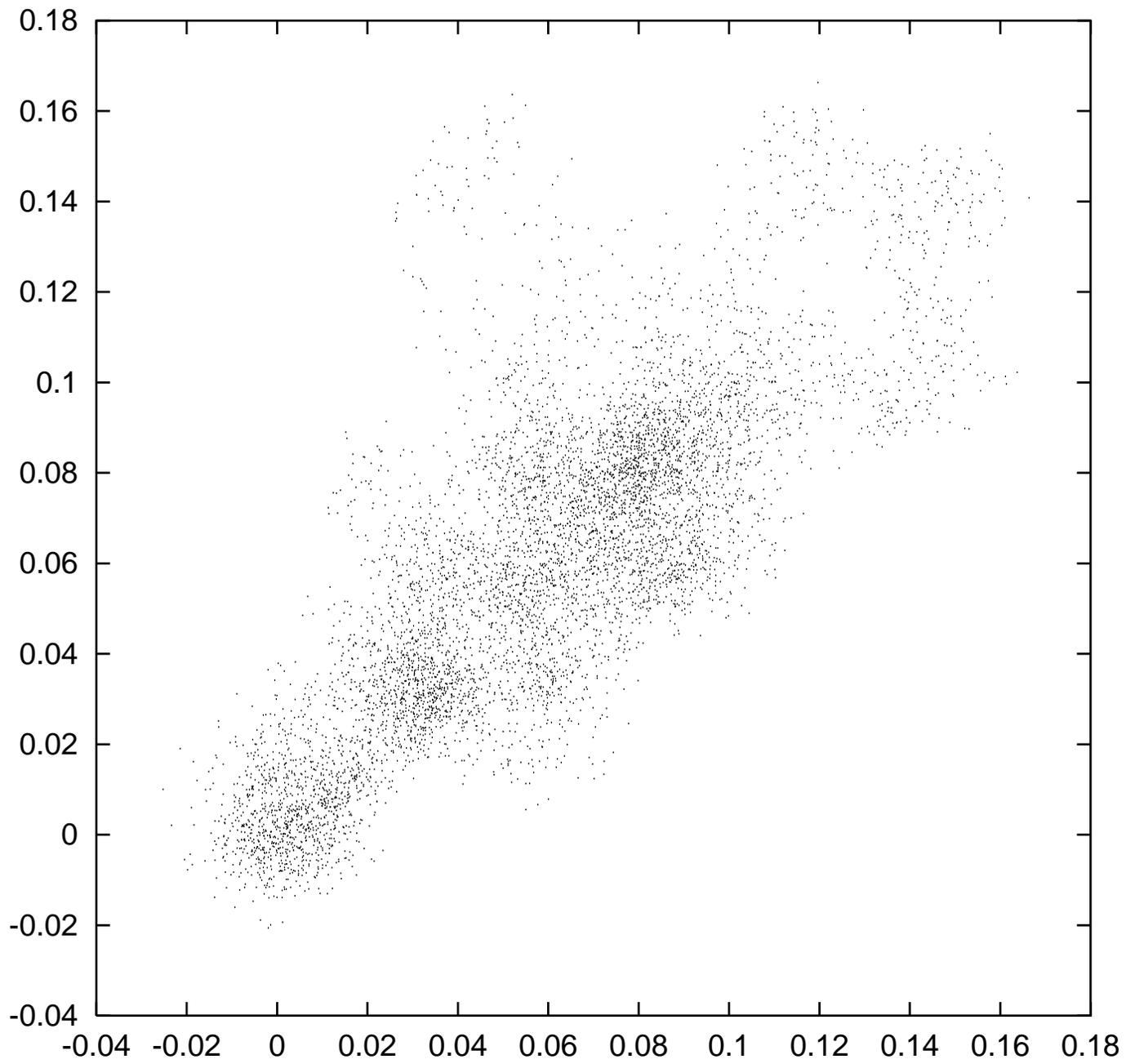
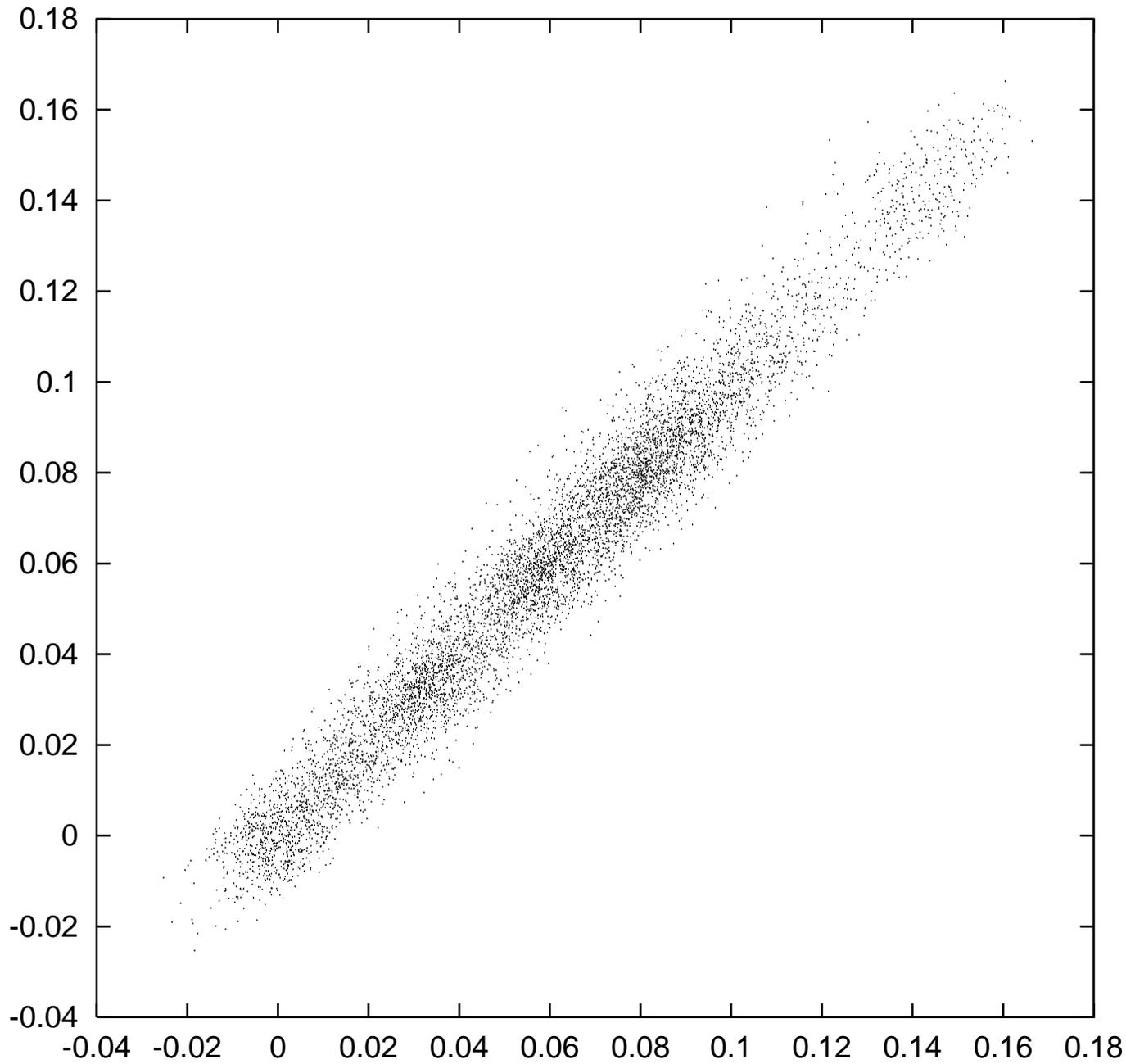
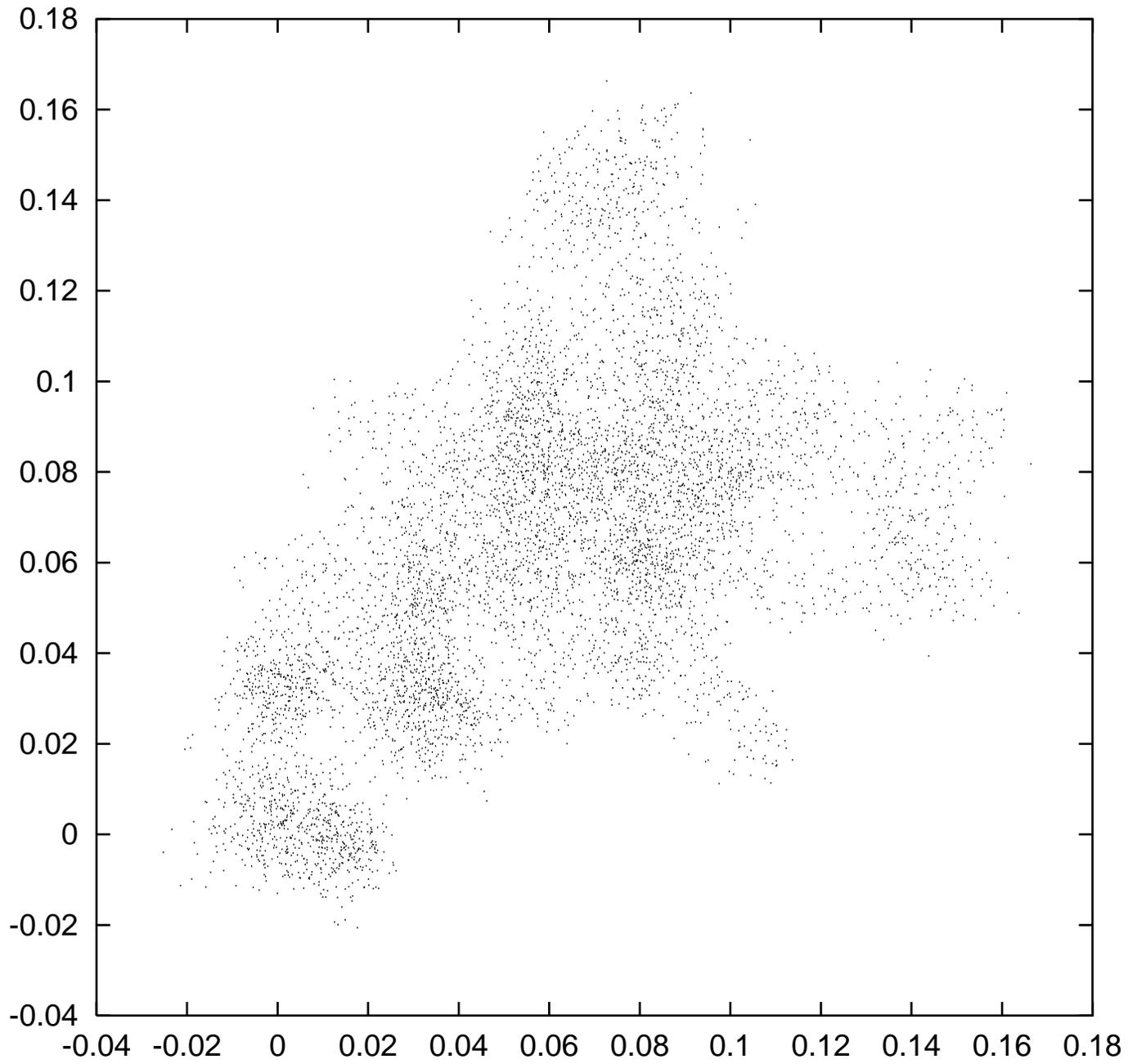
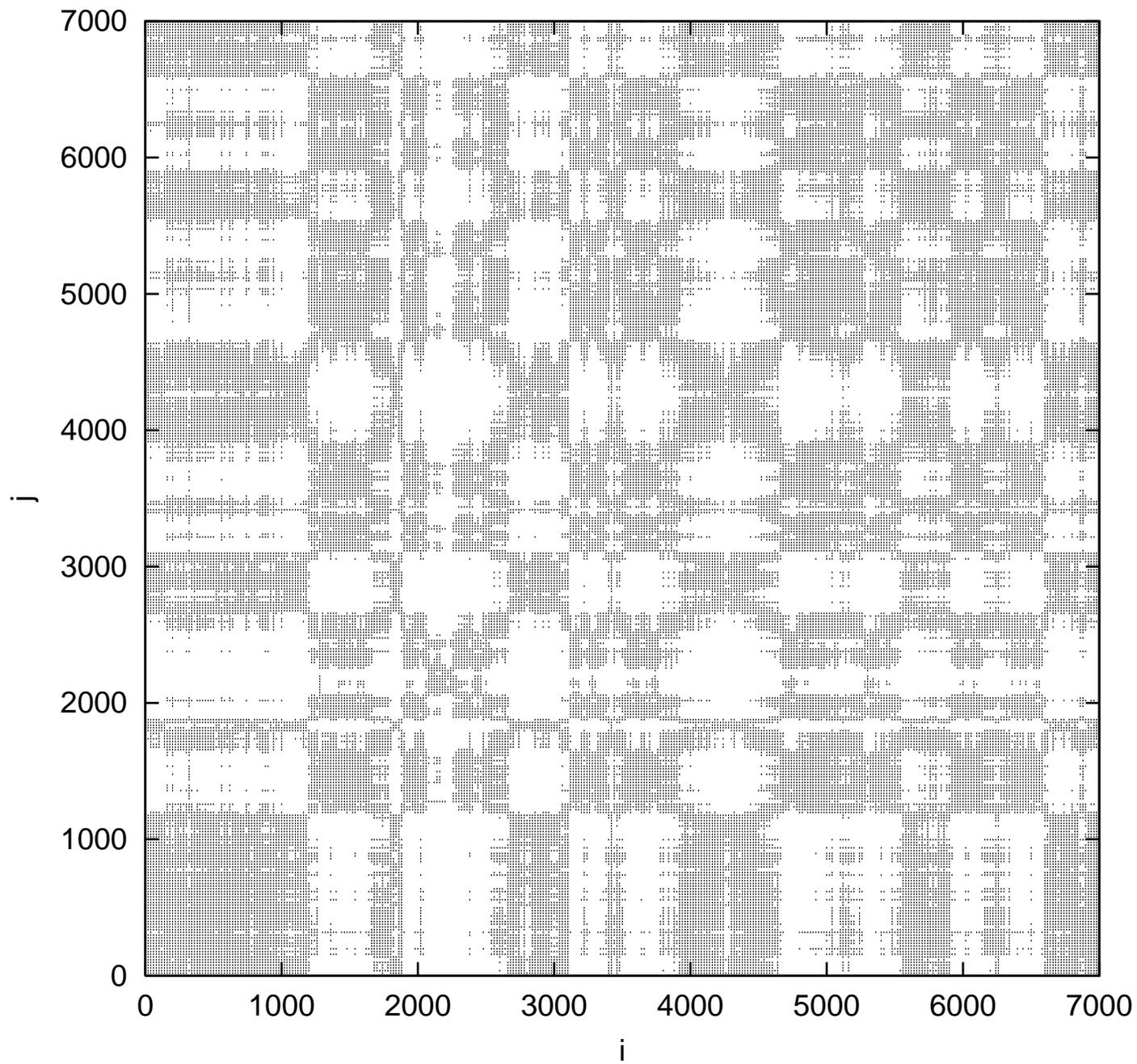
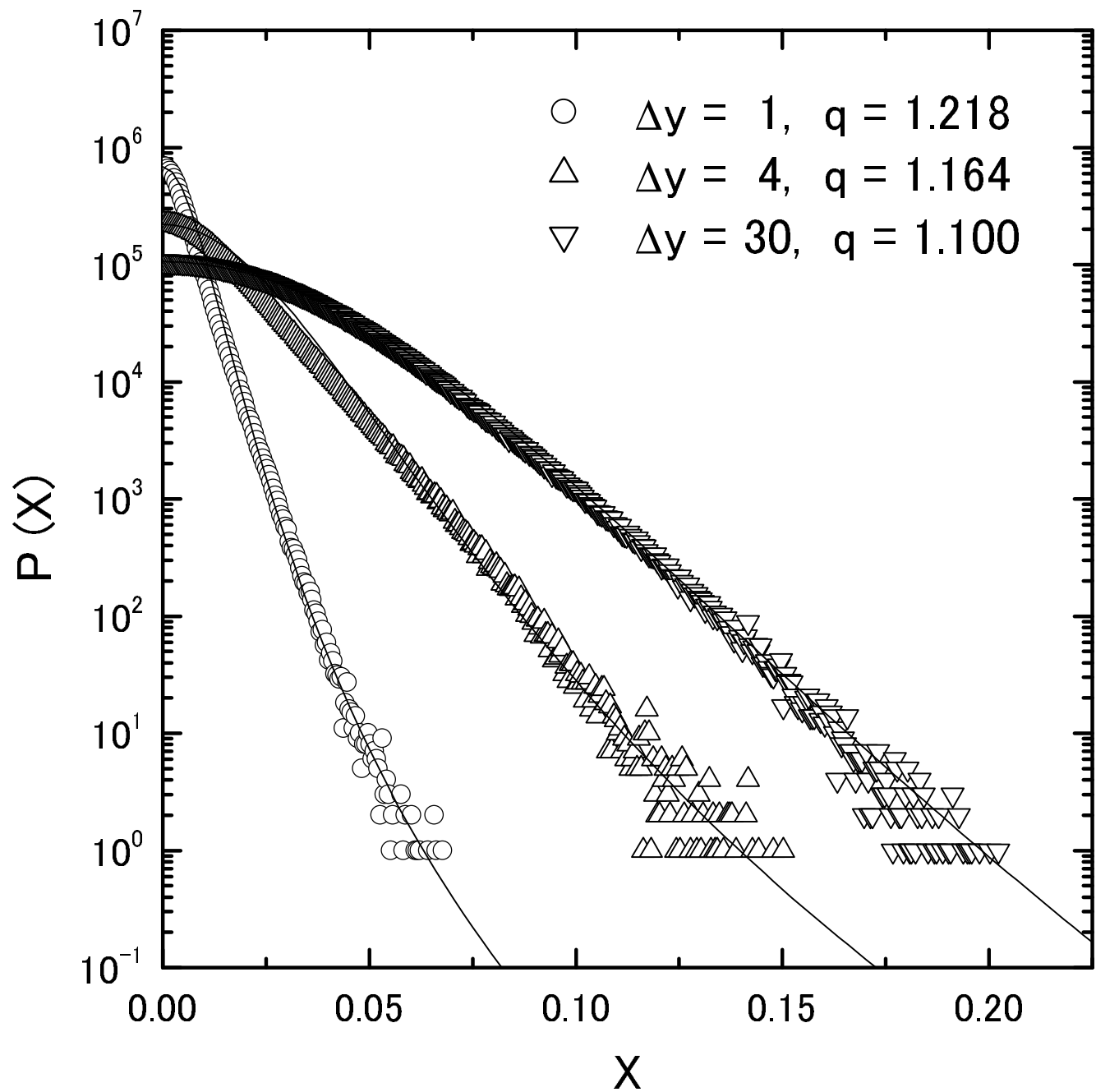
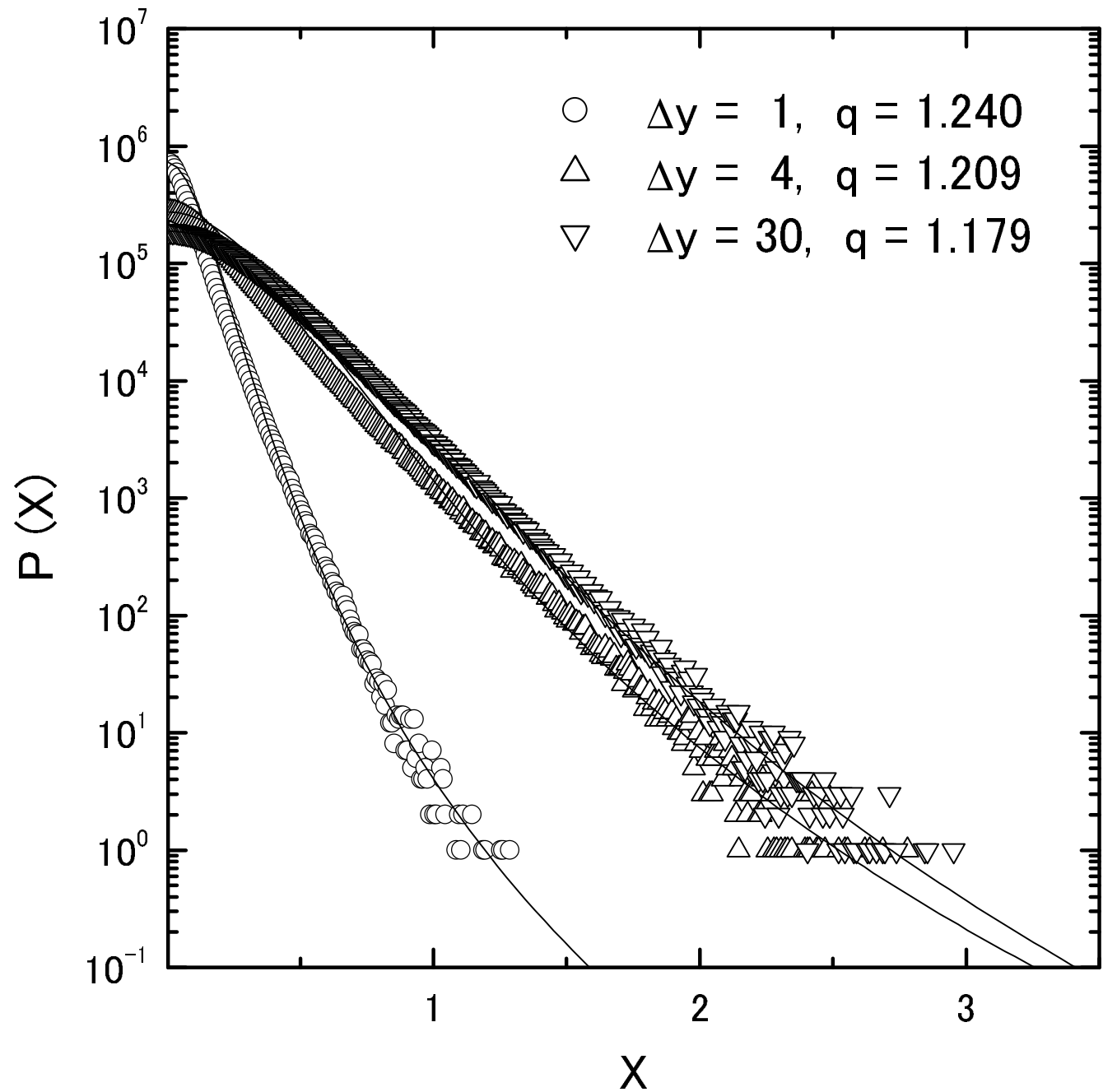
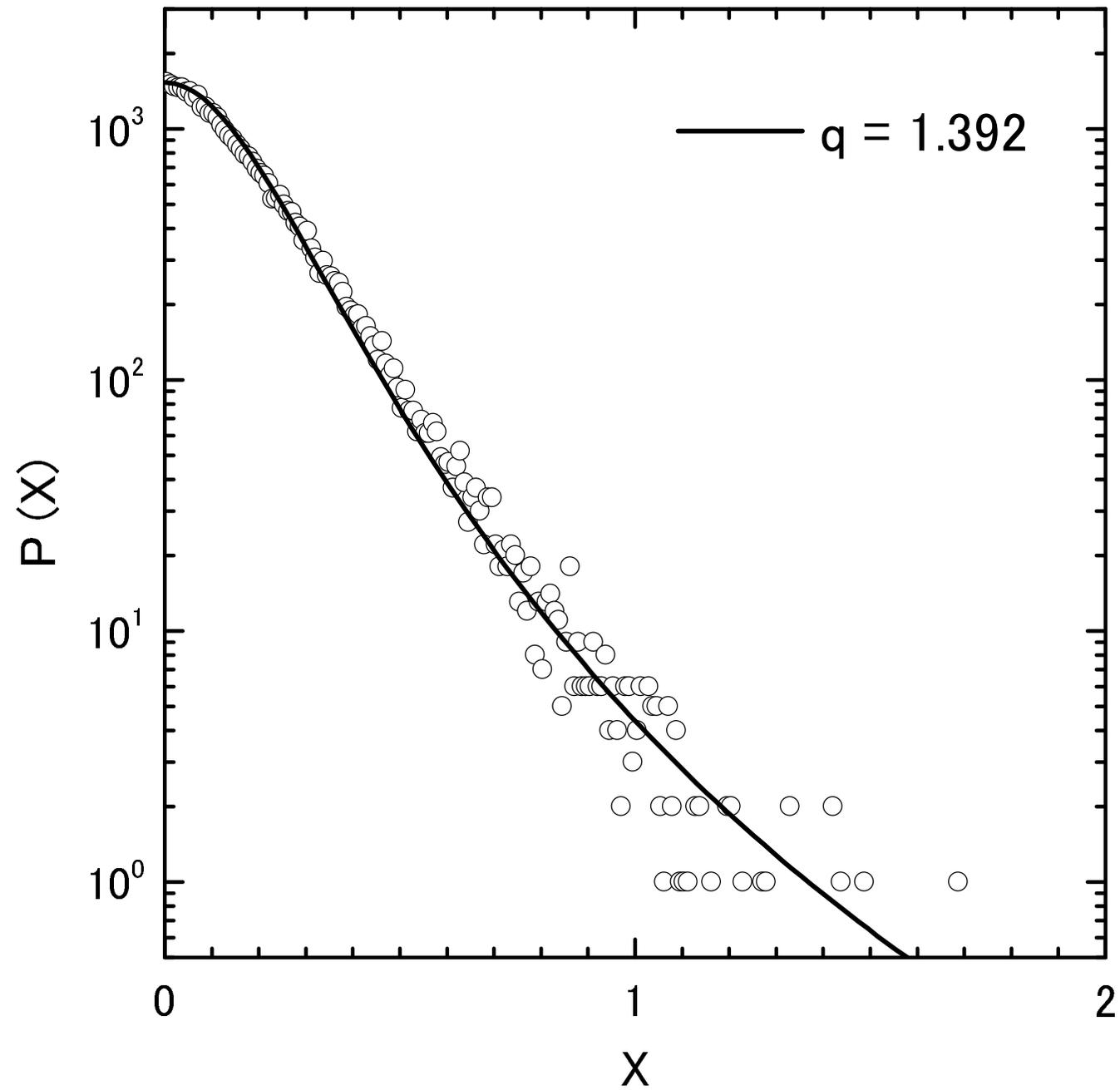
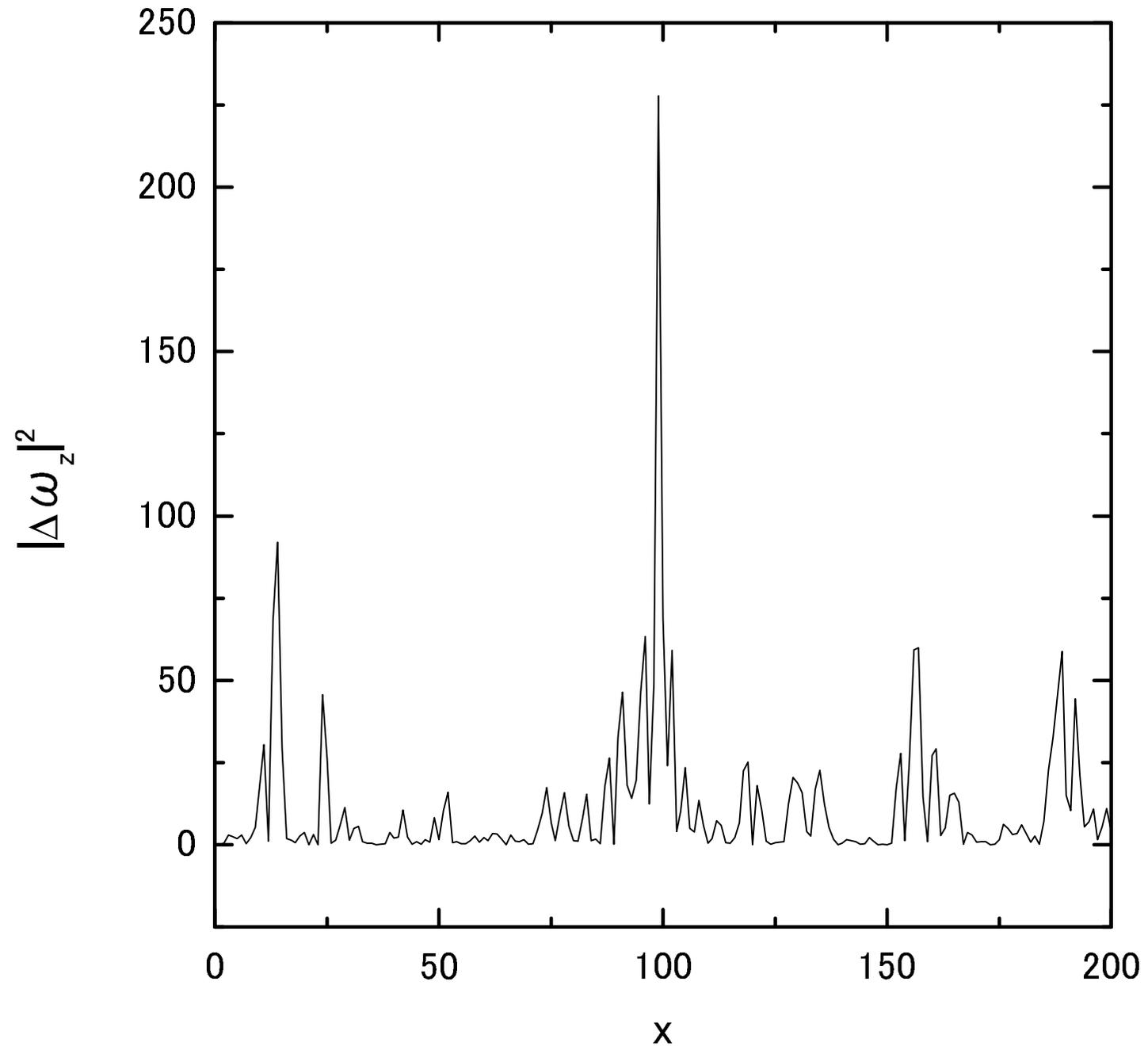
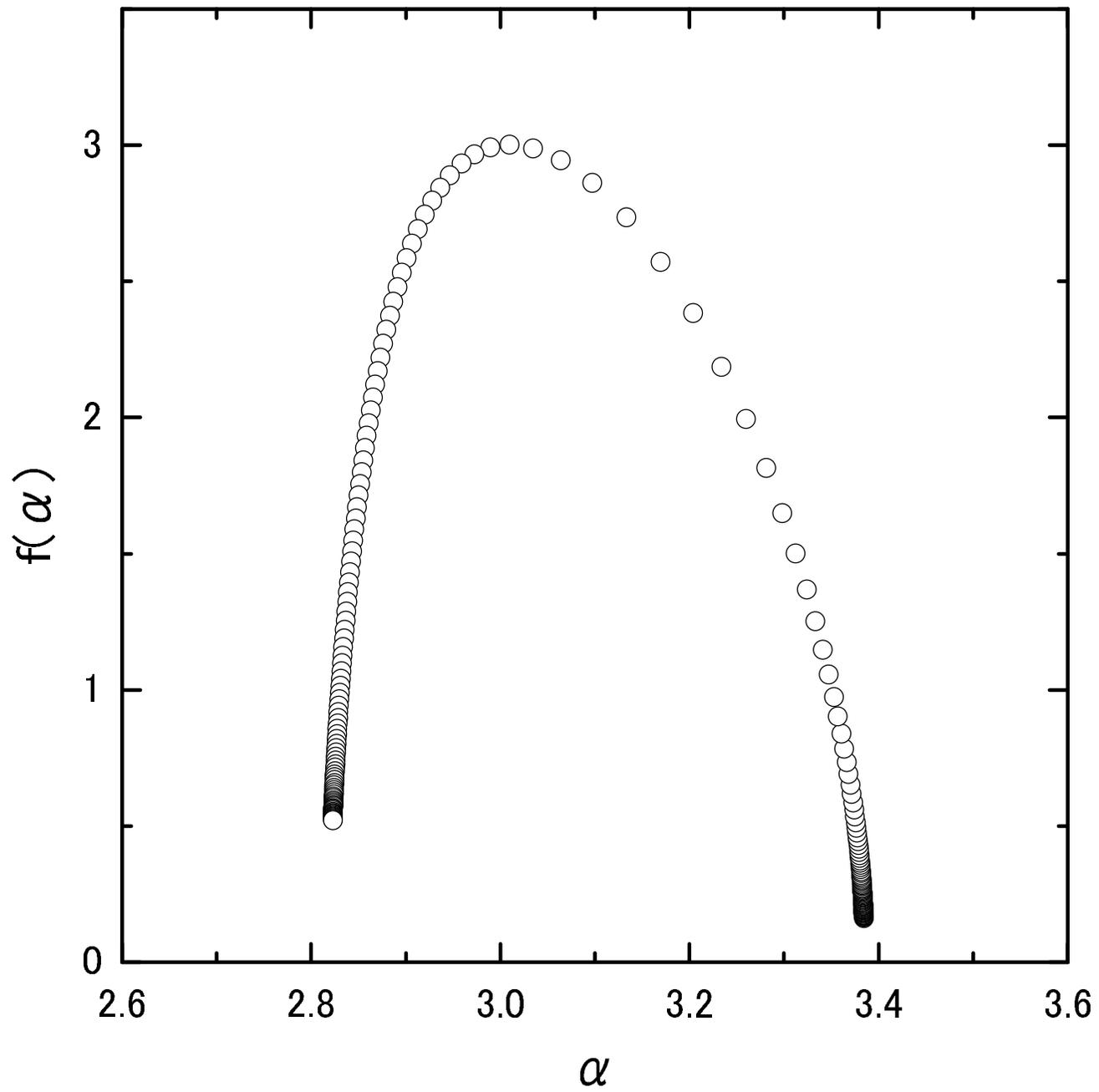
\end{document}